\documentclass[12pt]{iopart}
\usepackage{graphicx}
\catcode`\@=11
\def\numberbysection{\@addtoreset{equation}{section}
\def\theequation{\thesection.\arabic{equation}}}

\newcommand{\bea}{\begin{eqnarray*}}
\newcommand{\beq}{\begin{equation}}
\newcommand{\beqa}{\begin{eqnarray}}

\newcommand{\eea}{\end{eqnarray*}}
\newcommand{\eeq}{\end{equation}}
\newcommand{\eeqa}{\end{eqnarray}}

\renewcommand{\d}{{\rm d}}
\newcommand{\eq}{{\rm eq}}

\def\stackunder#1#2{\mathrel{\mathop{#2}\limits_{#1}}}

\begin{document}
\title{Dynamics of condensation in zero-range processes}
\author{C. Godr\`{e}che\dag}

\address{\dag\ Service de Physique de l'\'Etat Condens\'e,
CEA Saclay, 91191 Gif-sur-Yvette cedex, France}

\begin{abstract}
The dynamics of a class of zero-range processes exhibiting a condensation
transition in the stationary state is studied.
The system evolves in time starting from a random
disordered initial condition.
The analytical study of the large-time
behaviour of the system in its mean-field geometry provides a guide for the
numerical study of the one-dimensional version of the model.
Most qualitative features of the mean-field case are still present in the
one-dimensional system, both in the condensed phase and at criticality.
In particular the scaling analysis, valid for the mean-field system
at large time and for large values of the site occupancy, still holds
in one dimension.
The dynamical exponent $z$, characteristic of the growth of the condensate,
is changed from its mean-field value $2$ to $3$.
In presence of a bias, the mean-field value $z=2$ is recovered.
The dynamical exponent $z_c$,
characteristic of the growth of critical fluctuations, is changed
from its mean-field value $2$ to a larger value, $z_c\simeq 5$.
In presence of a bias, $z_c\simeq 3$.

\end{abstract}


\submitto{\JPA}

\section{Introduction}

In a recent work, Kafri et al. \cite{wis1} gave a criterion for the
existence of phase separation in driven one-dimensional systems.
Discussions on this theme
have been the subject of a long series of investigations
\cite{sven,lahir1,evans1,evans2,ritt1,ritt2,
korn,lahir2,das,sasa1,sasa2,mett,evans2002,barma,manoj}.
The criterion relies on a mapping, in the stationary state,
of the system onto a zero-range process.

For example, consider a two-species model, where each site of a
one-dimensional ring is either occupied by a ($+$) particle, or by a ($-$)
particle, or is vacant ($0$). Positive particles are driven to the right,
negative particles to the left, and they exchange positions when meeting.
Define domains as uninterrupted sequences of ($+)$ or ($-)$ particles. Then
identify the zeros with boxes, the particles with balls, and a domain with
the content of the box attached to the zero to its left, say. Then the
system of ($\pm )$ particles can be viewed as a system of balls in boxes,
where the rate at which a ball leaves a box is given by the current out of a
domain. This process is known as a zero-range process (ZRP) \cite{zrp1,zrp2}%
, more precisely defined below. Note that such a mapping of the particle
system to the corresponding ZRP gives a coarse-grained description of the
former, because the spatial structure inside domains is lost.
Once the hopping rate $d_{k}$ out of a box containing $k$ balls is known,
the analysis of the stationary state of the ZRP is straightforward.
Depending on the behaviour of $d_{k}$ at large $k$, it is easy to show that
there is, or there is not, a condensation transition, where a macroscopic
fraction of balls is contained in one box, the condensate. This corresponds
in the original particle system to a condensed phase where a single domain
contains a macroscopic fraction of the particles.

This is, in essence, the analysis given in~\cite{wis1}.
Examples of particle systems with two species satisfying the criterion for the
existence of phase separation have been subsequently given in ~\cite{wis2}.

These considerations concern the stationary state. We are thus naturally
lead to address the question of the dynamics, both for the underlying
original microscopic particle models, and for their coarse-grained counterpart,
the ZRP.

For the latter, the agenda is clear. The question is to determine the
temporal evolution of the system starting from a random initial condition,
once given the hopping rate out of a box.
The present paper is devoted to the study of the dynamics of a class of ZRP
giving rise to a condensation transition in their stationary state. We will
proceed in two steps. We study first the system in its mean-field geometry,
where all boxes are connected. This gives then a framework for the study of
the one-dimensional system.

The dynamics of the microscopic models
defined in ~\cite{wis2},
or more generally of two-species models satisfying the
criterion for phase separation,
is more subtle because it involves two
(a priori different) scales of time: the intra-domain dynamics,
corresponding to the equilibration inside a domain, and the inter-domain
dynamics, corresponding to the coarsening of domains. One can therefore
wonder whether during its temporal evolution the particle system is
equivalent to a ZRP with constant hopping rate. This study is postponed to
another publication \cite{ustocome}.

The aim of this paper is therefore two-fold. Firstly it addresses the
question of the dynamics of zero-range processes \textit{per se}. Indeed,
ZRP are sufficiently ubiquitous in the field of stochastic processes to
warrant the interest of such a study. For instance, the analysis of the zeta
urn model~\cite{zeta1,zeta2} is surprisingly close to that
presented in the next sections. Secondly it provides a template for the
study of the dynamics of particle systems, whenever these can be mapped, at
least at a coarse-grained level, to a ZRP, as explained above.

\section{ZRP: first definitions}

\subsection{Mean-field geometry and equilibrium: general formalism}

The zero-range process studied hereafter is defined as follows. Consider a
one-dimensional lattice of $M$ sites (or boxes), $i=1,\ldots ,M$. In each
box we have $N_{i}$ indistinguishable particles such that 
\[
\sum_{i=1}^{M}N_{i}=N. 
\]
The dynamics of the system is given by the rates at which a particle leaves
the departure box $d$, containing $N_{d}=k$ particles, and is put in the
arrival box $a$ containing $N_{a}=l$ particles. The hopping rate, denoted by
$W_{kl}$, a priori depends both on $k$ and $l$.

A configuration of the system is specified by the occupation numbers $%
N_{i}(t)$, hence a complete knowledge of its dynamics involves the
determination of $\mathcal{P}(N_{1,}N_{2},\ldots ,N_{M})$, the probability
of finding the system in a given configuration at time $t$. Hereafter we
will only focus our attention on the probability of finding $k$ particles in
the generic box $i=1$, 
\[
f_{k}(t)=\mathcal{P}(N_{1}(t)=k),
\]
that is, the marginal distribution of the occupation number $N_{1}(t)$ of
this box. Conservation of probability and of density yields 
\begin{eqnarray}
\sum_{k=0}^{\infty }f_{k}(t) &=&1,  \label{sumrule1} \\
\sum_{k=1}^{\infty }k\,f_{k}(t) &=&\rho ,  \label{sumrule2}
\end{eqnarray}
where we have taken the thermodynamic limit $N\rightarrow \infty
,M\rightarrow \infty $, with fixed density $\rho =N/M$.

Suppose now that the system has the mean-field geometry, i.e., all the boxes
are connected. In this case the temporal evolution of the occupation
probability $f_{k}(t)$ is explicitly given by the master equation 
\begin{eqnarray*}
\frac{\d f_{k}(t)}{\d t} &=&\mu _{k+1}\,f_{k+1}+\lambda _{k-1}\,f_{k-1}-(\mu
_{k}+\lambda _{k})f_{k}\qquad (k\geq 1), \\
\frac{\d f_{0}(t)}{\d t} &=&\mu _{1}\,f_{1}-\lambda _{0}f_{0},
\end{eqnarray*}
where
\begin{eqnarray}
\mu _{k} &=&\sum_{l=0}W_{kl}\,f_{l},  \label{mu} \\
\lambda _{k} &=&\sum_{l=1}W_{lk}\,f_{l},  \label{lam}
\end{eqnarray}
are respectively the rates at which a particle leaves box number 1
($N_{1}=k\rightarrow N_{1}=k-1$), or arrives in this box
($N_{1}=k\rightarrow N_{1}=k+1 $). This is the master equation for a random walk for $N_{1}$, i.e.,
on the positive integers $k=0,1,\ldots $. The equation for $f_{0}$ is
special because one cannot select an empty box as a departure box, nor can $%
N_{1}$ be negative. The conservation law $\d\sum k\,f_{k}/dt=0$, valid at all
times, implies the sum rule
\begin{equation}
\sum_{k=1}^{\infty }\mu _{k}\,f_{k}=\sum_{k=0}^{\infty }\lambda _{k}\,f_{k}.
\label{sumrule}
\end{equation}

In the stationary state ($\dot{f}_{k}=0$), we find 
\[
\frac{f_{k+1,\eq}}{f_{k,\eq}}=\frac{\lambda _{k}}{\mu _{k+1}}, 
\]
which expresses the detailed balance condition at equilibrium, yielding
\[
f_{k,\eq}=\frac{\lambda _{0}\ldots \lambda _{k-1}}{\mu _{1}\ldots \mu _{k}}%
f_{0,\eq},
\]
where $f_{0,\eq}$ is fixed by normalisation.

Hereafter we are interested in the case where the rate $W_{kl}$ only depends
on the occupation of the departure box. We denote the hopping rate out of
the departure box by $W_{kl}=d_{k}$. (Alternatively one can consider cases
where $W_{kl}$ only depends on the occupation of the arrival box, as for the
zeta urn model \cite{zeta1,zeta2}: $W_{kl}=a_{l}$ is the hopping rate \emph{%
into} the arrival box; see below.) We thus have 
\begin{equation}
\mu _{k}=d_{k},\qquad \lambda _{k}=\sum_{l=1}^{\infty }d_{l}f_{l}\equiv \bar{%
d}_{t}.  \label{rates}
\end{equation}
Note that (\ref{sumrule}) is automatically satisfied. We obtain
\begin{equation}
f_{k,\eq}=\frac{p_{k}z^{k}}{\sum_{k=0}^{\infty }p_{k}z^{k}},  \label{fkeq}
\end{equation}
where $z\equiv \sum d_{l}f_{l,\eq}=\bar{d}_{\eq}$, and 
\begin{eqnarray}
p_{k} &=&\frac{1}{d_{1}\ldots d_{k}}\qquad (k\geq 1)  \label{pk} \\
p_{0} &=&1.  \nonumber
\end{eqnarray}

It is important to emphasize here that the results above, equations (\ref
{fkeq}) and (\ref{pk}), also hold for the equivalent one-dimensional ZRP in
the stationary state.
Indeed a well-known property of ZRP states that in the
stationary state the probability $\mathcal{P}(N_{1,}N_{2},\ldots ,N_{M})$
is given by a product measure \cite{zrp1,zrp2}.
Explicitly $\mathcal{P}(N_{1,}N_{2},\ldots ,N_{M})$,
with $\sum N_i=N$, factorizes into the product
$f(N_{1})f(N_{2})\ldots f(N_{M})$, where each
factor $f(N_{i}=k)$ is equal to $f_{k,\eq}$ as given by
(\ref{fkeq}) and (\ref{pk}).
For an infinite system this factorization property holds
at all time in the mean field geometry, by essence.

\subsection{Class of ZRP with a condensation transition}

Let us apply this formalism to the case where, at large $k$%
\begin{equation}
d_{k}\approx 1+\frac{b}{k}+\cdots . \label{dkasympt}
\end{equation}
Then it is easy to see that, in the same regime, $p_{k}\sim k^{-b}$. Hence
the series $\sum_{k=0}^{\infty }p_{k}z^{k}$ has a radius of convergence $%
z_{c}=1$. The unknown parameter $z\equiv \bar{d}_{\eq}$ is determined by
using (\ref{sumrule2}), which yields
\begin{equation}
\sum_{k=1}^{\infty }k\,f_{k,\eq}=\frac{\sum_{k=0}^{\infty }k\,p_{k}z^{k}}{%
\sum_{k=0}^{\infty }p_{k}z^{k}}=\rho .  \label{col}
\end{equation}
This expression is monotonously increasing with $z$. When $z$ reaches the
value $z_{c}=1$, the right side is converging only if $b>2$. Therefore we
have the following two cases.

$\bullet $ If $b<2$, equation (\ref{col}) possesses a solution $z(\rho )$
for any value of $\rho $. The occupation probabilities are exponentially
decaying as 
\begin{equation}
f_{k,\eq}=\frac{p_{k}z^{k}}{\sum_{k=0}^{\infty }p_{k}z^{k}}\sim k^{-b}\exp
\left( -\frac{k}{\xi }\right) \qquad (\mathrm{fluid})  \label{fluid},
\end{equation}
where the correlation length $\xi =\left|\ln z\right|^{-1} $.
The system is in a
fluid phase.

$\bullet $ If $b>2$, there exists a critical value of the density, where the
correlation length diverges, attained when $z=z_{c}=1$:
\begin{equation}
\rho _{c}=\frac{\sum_{k=0}^{\infty }k\,p_{k}}{\sum_{k=0}^{\infty }p_{k}}.
\label{rocrit}
\end{equation}
This is the equation of a transition line. In other
terms, if $\rho <\rho _{c}$, then (\ref{col}) has a solution $z(\rho )$, in
which case the system is in the fluid phase described above (see equation (%
\ref{fluid})).
Otherwise if $\rho >\rho _{c}$, then necessarily $z\equiv
\bar{d}_{\eq}=1$.
The system is composed of a critical part and a condensate.
The critical part is described by the algebraically decaying distribution 
\begin{equation}
f_{k,\eq}=\frac{p_{k}}{\sum_{k=0}^{\infty }p_{k}}\sim k^{-b}\qquad (\mathrm{%
critical})  \label{fkeqcrit}
\end{equation}
contributing to $\rho _{c}$, as shown by (\ref{rocrit}). The condensate,
contributing to the remaining $\rho -\rho _{c}$, corresponds to particles
sitting in a single box.
At criticality ($\rho
=\rho _{c}$) the condensate disappears and (\ref{fkeqcrit}) still holds.

Let us illustrate the above formalism for the explicit choice of hopping
rate 
\begin{equation}
d_{k}=1+\frac{b}{k},  \label{dk}
\end{equation}
which is used in the numerical studies of the forthcoming sections. We have 
\begin{equation}
f_{k,\eq}=(b-1)\frac{\Gamma (b)\,\Gamma (k+1)}{\Gamma (k+b+1)}\stackunder{%
k\rightarrow \infty }{\approx }(b-1)\Gamma (b)\,k^{-b},  \label{fkeqcrit_2}
\end{equation}
and
\begin{equation}
\rho _{c}=\frac{1}{b-2},  \label{roc}
\end{equation}
which is the equation of the critical line in the $b-\rho $ plane.
It is simple to check using (\ref{fkeqcrit_2}) that we have consistently $%
\bar{d}_{\eq}=1$, both at criticality and in the condensed phase.

\section{ZRP: dynamics in the mean-field geometry}

From now on, we consider a hopping rate $d_{k}$ out of the departure box of
the form (\ref{dkasympt}), i.e., such as $d_{k}\approx 1+b/k$ at large $k$, and
for the sake of analytical or numerical illustrations we will choose the
explicit form (\ref{dk}), unless specified.

For an infinite system in the mean-field geometry, the knowledge of the
occupation probabilities $f_{k}(t)$ provides a complete description of its
dynamics.
We consider a system with Poissonian initial distribution of occupation
probabilities,
\[
f_{k}(0)=\e^{-\rho }\frac{\rho ^{k}}{k!},
\]
i.e., such that initially particles are distributed at random amongst boxes.
The temporal
evolution of the occupation probabilities $f_{k}$ is given by the master
equation
\begin{eqnarray}
\frac{\d f_{k}(t)}{\d t} &=&d_{k+1}\,f_{k+1}+\bar{d}_{t}\,f_{k-1}-(d_{k}+\bar{d}%
_{t})f_{k}\qquad (k\geq 1),  \label{master} \\
\frac{\d f_{0}(t)}{\d t} &=&d_{1}\,f_{1}-\bar{d}_{t}f_{0}.
\nonumber
\end{eqnarray}
This set of equations is non linear because $\bar{d}_{t}\equiv \sum
d_{l}f_{l}$ is itself a function of the $f_{k}(t)$. Hence there is no
explicit solution of these equations in closed form.
Yet one can extract
from them an analytical description of the dynamics of the system
at long times, both in the condensed phase, and at criticality.
The structure of the reasoning
borrows to former studies on urn models \cite{zeta1,zeta2}.
(For a review, see \cite{revue}.)

As we show below, there exists two different regimes in the evolution of the
system, both in the condensed phase or at criticality, which we study
successively.

\subsection{Condensed phase ($\rho >\rho _{c}$)}

Since $\bar{d}_{\eq}=1$, we set, for large times, 
\begin{equation}
\bar{d}_{t}\approx 1+A\,\,\varepsilon _{t},
\label{dbart}
\end{equation}
where the small time scale $\varepsilon _{t}$ is to be determined. As shown
below, $\varepsilon _{t}\sim t^{-\frac{1}{2}}$.

\paragraph{Regime I: $k$ fixed, $t$ large.}

For $t$ large enough, boxes empty ($d_{k}$) faster than they fill ($\bar{d}%
_{t}$). In this regime there is convergence to equilibrium, hence we set 
\begin{equation}
f_{k}(t)\approx f_{k,\eq}(1+v_{k}\,\varepsilon _{t}),  \label{fkreg1}
\end{equation}
with $f_{k,\eq}$ given by (\ref{fkeqcrit}), and where the $v_{k}$ are
unknown. This expression carried into (\ref{master}) yields the stationary
equation $\dot{f}_{k}=0$, because the derivative $\dot{f}_{k}$, proportional
to $\dot{\varepsilon}_{t}$, is negligible compared to the right-hand side. We
thus obtain the detailed balance condition:
\[
\frac{f_{k+1,\eq}}{f_{k,\eq}}\frac{1+v_{k+1}\,\varepsilon _{t}}{%
1+v_{k}\,\varepsilon _{t}}=\frac{1+A\,\varepsilon _{t}}{d_{k+1}}.
\]
Using (\ref{fkeqcrit}) and (\ref{pk}), we obtain, at leading order in $%
\,\varepsilon _{t}$, $v_{k+1}-v_{k}=A$, and finally 
\begin{equation}
v_{k}=v_{0}+k\,A.  \label{vk}
\end{equation}
At this stage, $v_{0}$ and the amplitude $A$ are still to be determined.

\paragraph{Regime II: $k$ and $t$ are simultaneously large.}

This is the scaling regime, with scaling variable $u=k\,\varepsilon _{t}$.
We look for a similarity solution of (\ref{master}) of the form 
\begin{equation}
f_{k}(t)\approx \varepsilon _{t}^{2}g(u).  \label{fkscalcond}
\end{equation}
Following the treatment of \cite{zeta1} we obtain for $g(u)$ the linear
differential equation 
\[
g^{\prime \prime }(u)+\left( \frac{u}{2}-A+\frac{b}{u}\right) g^{\prime
}(u)+\left( 1-\frac{b}{u^{2}}\right) g(u)=0.
\]
This is precisely the differential equation found in \cite{zeta1,zeta2}.
Moreover this method leads to $\varepsilon_{t}\sim t^{-\frac{1}{2}}$
\cite{zeta1}.
The amplitude $A$ can be determined by the fact that the equation has an
acceptable solution $g(u)$ vanishing as $u\rightarrow 0$ and $u\rightarrow
\infty $ \cite{zeta1}. It is a universal quantity, only depending on the
value of $b$. The normalisation of the solution is fixed by the sum rule (%
\ref{sumrule1}) yielding \cite{zeta1} 
\begin{equation}
\int_{0}^{\infty }\d u\,ug(u)=\rho -\rho _{c}.  \label{rs1_cond}
\end{equation}
The sum rule (\ref{sumrule2}) leads to \cite{zeta1}
\begin{equation}
v_{0}+A\rho _{c}=-\int_{0}^{\infty }\d u\,g(u).  \label{rs2_cond}
\end{equation}
The differential equation above has no closed form solution. However further
information on the form of the solution $g(u)$ can be found in \cite
{zeta1,zeta2}. In particular $g(u)/(\rho -\rho _{c})$ is universal
in the sense that it only depends on the parameter $b$
(see the discussion below).

An intuitive description of the dynamics of the condensate
in the scaling regime is as follows.
The typical occupancy $k_{\rm cond}$ of the boxes making
the condensate scales as $t^{\frac{1}{2}}$.
The total number of
particles in the condensate is equal to $M(\rho -\rho _{c})$,
the remaining $M\,\rho _{c}$ lying in the fluid.
Therefore the number of boxes belonging
to the condensate scales as
$M (\rho -\rho _{c})t^{-\frac{1}{2}}$.

\subsection{Criticality ($\rho =\rho _{c}$)}

The analysis follows closely that done in \cite{zeta2}. We just sketch the
method here.

We set
\[
\bar{d}_{t}\approx 1+A\,\varepsilon _{t} ,
\]
with $\varepsilon _{t}=t^{-\omega }$, where the exponent $\omega $ is to be
determined, and we consider the same two regimes as above.
In regime I, we still set (\ref{fkreg1}) for $f_{k}(t)$.
The reasoning
leading to the relationship $v_{k}=v_{0}+k\,A$ (see (\ref{vk}))
is still valid here.
In regime II, we look for a similarity solution to (\ref{master}) of the
form 
\begin{equation}
f_{k}(t)\approx f_{k,\eq}\,g_c(u)\qquad u=k\,t^{-\frac{1}{2}}.
\label{fkscal_crit}
\end{equation}
The sum rules (\ref{sumrule1}) and (\ref{sumrule2}) lead respectively to
the following equations, provided that $b>3$,
\begin{eqnarray}
v_{0}+A\rho _{c} &=&0,  \label{v0_plus_aroc} \\
t^{-\omega }\left( v_{0}\rho _{c}+A\mu _{c}\right)
&=&t^{-(b-2)/2}(b-1)\Gamma (b)\int_{0}^{\infty }\d u\,u^{1-b}(1-g_c(u)),
\label{v0roc}
\end{eqnarray}
where, using (\ref{fkeqcrit_2}), $\mu _{c}=\sum
k^{2}\,f_{k,\eq}=(b+1)(b-2)^{-1}(b-3)^{-1}$.
Equation (\ref{v0roc}) fixes the value of $%
\omega $: 
\begin{equation}
\omega =(b-2)/2.  \label{omega}
\end{equation}
The differential equation obeyed by $g_c(u)$ is obtained by carrying (\ref
{fkscal_crit}) into (\ref{master}).
It reads

\[
g_c^{\prime \prime }(u)+\left( \frac{u}{2}-\frac{b}{u}\right) g_c^{\prime
}(u)=0, 
\]
the solution of which is, with $g_c(0)=1$,
\begin{equation}
g_c(u)=\frac{2^{-b}}{\Gamma (\frac{b+1}{2})}\int_{u}^{\infty
}\d y\,y^{b}\e^{-y^{2}/4}.
\label{gu_anal}
\end{equation}
The fall-off of $g_c(u)$
for $u\gg 1$ is very fast: $g_c(u)\sim \exp (-u^{2}/4)$,
hence $f_{k}(t)\sim \exp (-k^{2}/4t)$.
We finally obtain
\[
A=\frac{(b-1)\Gamma (b)}{\mu _{c}-\rho _{c}^{2}}\int_{0}^{\infty
}\d u\,u^{1-b}(1-g_c(u))
=\frac{(b-2)(b-3)}{b-1}\Gamma \left( \frac{b}{2}\right)
. 
\]

\subsection{Universality}

To summarize, for any hopping rate of the form $d_{k}\approx 1+b/k$, the
scaling functions, $g(u)$ in the condensed phase (more
precisely: $g(u)/(\rho -\rho _{c})$),
and $g_c(u)$ at criticality, are universal.
In both cases the scaling variable is $u=k\,t^{-1/2}$.
The critical density $\rho _{c}$,
and, as a consequence, any quantity depending on $\rho_c$, such as
the amplitude $v_{0}$,
are non universal,
with values depending on the precise definition of
$d_{k}$.
As noted above, the amplitude $A$ is a universal quantity in the
condensed phase.

Universality can be checked by comparing the results obtained by
a numerical integration of the master equation (\ref{master}),
for different choices of the rate $d_{k}$.
Besides the form $d_{k}=1+b/k$ of eq. (\ref{dk})
we investigated the following two other expressions
\begin{eqnarray}
d_{k} &=&1+\frac{b}{k}+\frac{c}{k^{2}},  \label{other1} \\
d_{k} &=&\left( 1+\frac{1}{k}\right) ^{b}.  \label{other2}
\end{eqnarray}

\begin{figure}[htb]
\begin{center}
\includegraphics[angle=0,width=.7\linewidth]{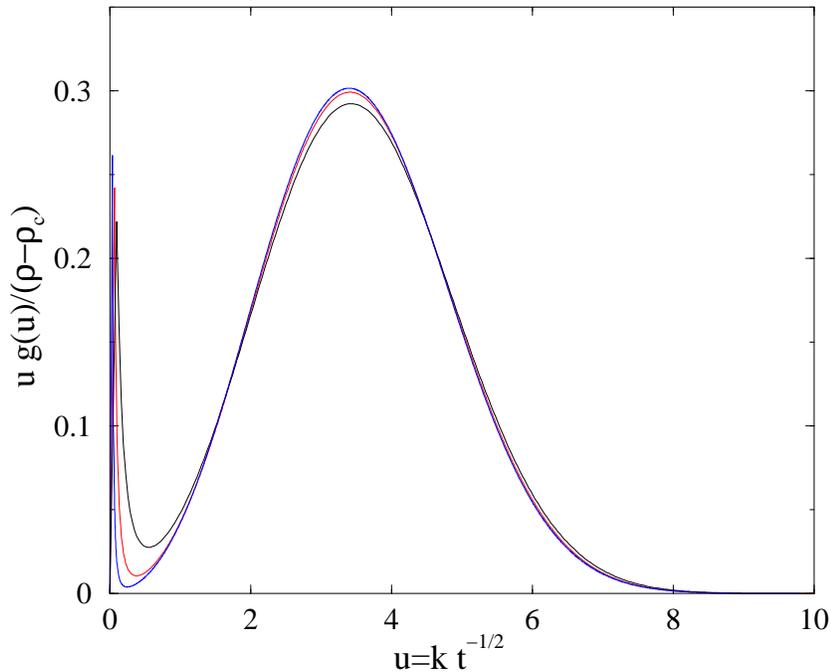}
\caption{\small
Mean-field geometry: normalised scaling function $u g(u)/(\rho-\rho_c)$
in the condensed phase, for times equal to $10^3$, $10^4$, and $10^5$.
($b=4$, $\rho=20\rho_c=10$.)
}
\label{f1}
\end{center}
\end{figure}

\begin{figure}[htb]
\begin{center}
\includegraphics[angle=0,width=.7\linewidth]{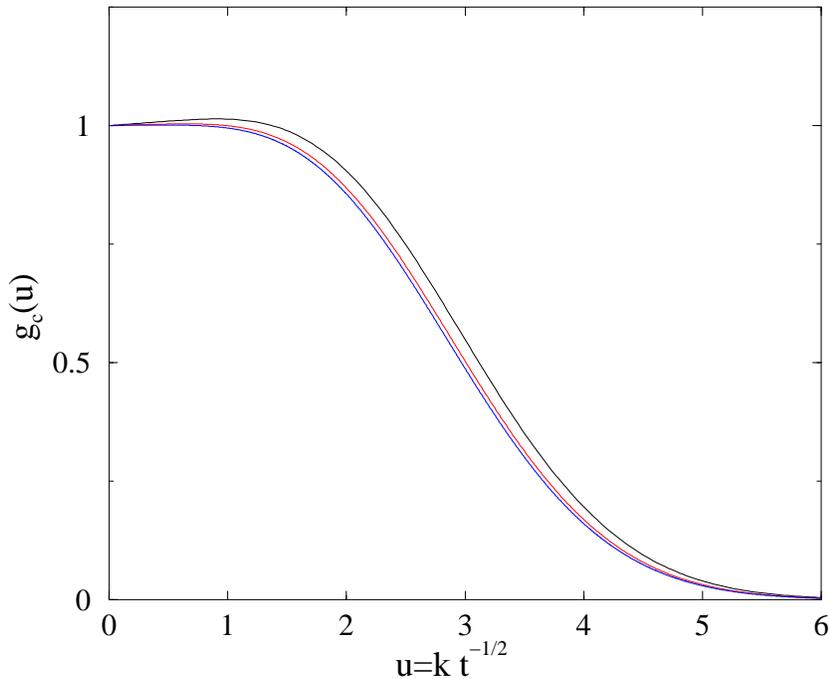}
\caption{\small
Mean-field geometry: scaling function $g_c(u)$ at criticality
for times equal to $10^3$, $10^4$, and $10^5$.
($b=4$, $\rho=\rho_c=0.5$.)
}
\label{f2}
\end{center}
\end{figure}

Figure~\ref{f1} depicts the normalized scaling function
$u g(u)/(\rho -\rho _{c})$
in the condensed phase, for times equal to $10^3$, $10^4$, and $10^5$.
It was obtained with the choice of rate (\ref{dk}),
with $b=4$, and for a density $\rho =20\rho _{c}$.
The curves obtained with the choice (\ref
{other1}) and $c=10$, or with (\ref{other2}), are hardly distinguishable
from the former one.
Figure~\ref{f2} depicts $g_c(u)$ at criticality, for the same range of times,
with the choice (\ref{dk}), and $b=4$.
Again, the curves obtained with the choice (\ref
{other1}), with $c=10$, or with (\ref{other2}),
are indistinguishable from the former one.
The limiting curve of figure~\ref{f2} coincides exactly
with the form (\ref{gu_anal}).

One also finds, for $\rho>\rho_c$ and $b=4$, $A\simeq 1.9$,
and $\int \d u g(u)/(\rho-\rho_c)\simeq 0.336$,
which are universal quantities.
Correspondingly, for $d_k$ given by (\ref{dk}),
and $\rho=20\rho_c$, $v_0\simeq -4.15$.


It is remarkable that the universality of the scaling
functions $g(u)$ and $g_c(u)$ extends
beyond the case of the ZRP under study. Indeed, for the zeta urn model,
defined by a rate $W_{kl}$ only depending on the occupation $l$ of the
arrival box, much of the formalism above is the same \cite{zeta1,zeta2}. An
intuitive explanation of this fact comes from the comparison of the rates of
the two models, at large values of time and occupation numbers. Consider the
rate $\mu _{k}$ and $\lambda _{k}$ defined in equations (\ref{mu}) and
(\ref{lam}). For the ZRP studied here these rates
are given by (cf. (\ref{rates}))

\[
\mu _{k}=d_{k}\approx 1+\frac{b}{k},\qquad \lambda _{k}
=\sum d_{l}f_{l}\equiv
\bar{d}_{t}\approx 1+A\varepsilon _{t},\qquad (\mathrm{ZRP})
\]
For the zeta urn model (see \cite{zeta1,zeta2} for detailed definitions), we
have 
\[
\mu _{k}=1,\qquad \lambda _{k}\approx \left( 1-\frac{\beta}{k}\right) \left(
1+A\varepsilon _{t}\right) \qquad (\mathrm{zeta}),
\]
where $\beta$ is the inverse temperature, playing the role of
the parameter $b$ in the model.
Simple inspection of the last two equations demonstrates the formal analogy
between the two models. This analogy is still stronger if one chooses the
rate $d_{k}$ given by (\ref{other2}). Then, at equilibrium, the formalisms
for the corresponding  ZRP and for the zeta urn model coincide.

\section{ZRP: dynamics in one dimension}

The framework of analytic results obtained in the mean-field case provides a
guide for the numerical study of the one-dimensional system. It turns out
that most of the features of the former survive in one dimension. In
particular the two regimes of time described in the previous section are
still present in the one-dimensional case. The main difference lies in the
values of the dynamical exponent $z$ characteristic of the growth of the
condensate, and of the critical dynamical exponent $z_{c}$ characteristic of
the growth of critical fluctuations.
In the condensed phase ($\rho >\rho_c$), numerical simulations, presented
hereafter, give evidence for $z = 3$, instead of $z = 2$
in the mean-field geometry.
At criticality $z_c\simeq5$ instead of $z_c = 2$, in the mean-field geometry.

We consider a system on a ring. Sites have multiple occupancy, given by the
occupation numbers $N_{i}(t)$, where $i=1,\ldots ,M$. At each time step a
site $i$ is chosen at random and, if not empty, a particle of this site is
transferred to one of the two neighbouring site, to the right or to the left
with equal probability, with a rate $d_{k}$ only depending on the number of
particles $N_{i}=k$, present on site $i$. As above, we consider a hopping
rate such that $d_{k}\sim 1+b/k$ at large $k$. The numerical simulations
described below were performed with the choice of rate given by (\ref{dk}).

\subsection{Condensed phase ($\rho >\rho _{c}$)}

We begin with the determination of the dynamical exponent $z$.
In regime II ($k$ and $t$ simultaneously large), let us
assume the scaling form, inspired by eq. (\ref{fkscalcond}),
\[
f_{k}(t)\approx \frac{1}{t^{2/z}}g\left( \frac{k}{t^{1/z}}\right) .
\]
Then, at large time, the mean squared occupation 
\begin{equation}
\mu _{2}=\left\langle N_{1}^{2}(t)\right\rangle =\sum_{k=1}^{\infty
}k^{2}\,f_{k}(t)  \label{mu2}
\end{equation}
should behave as 
\begin{equation}
\mu _{2}\sim t^{1/z}\int_{0}^{\infty }\d u\,u^{2}\,g(u).  \label{k2_cond}
\end{equation}
This quantity is easy to measure, and remains accurate
even for relatively large times.
More generally the moment of order $n$,
$\mu_n=\left\langle N_{1}^{n}(t)\right\rangle$,
scales as $t^{(1/z)(n-1)}$.
The behaviour (\ref{k2_cond}) is very well observed in numerical simulations,
with $z=3$ (see figure~\ref{f4} below).
Figure~\ref{f3} depicts the normalized scaling function
$ug(u)/(\rho -\rho _{c})$
for times equal to $10^3$, $3\,10^3$, and $10^4$,
with $u=kt^{-1/3}$.
The data are for $b=4$, and $\rho =20\,\rho _{c}=10$.

\begin{figure}[htb]
\begin{center}
\includegraphics[angle=0,width=.7\linewidth]{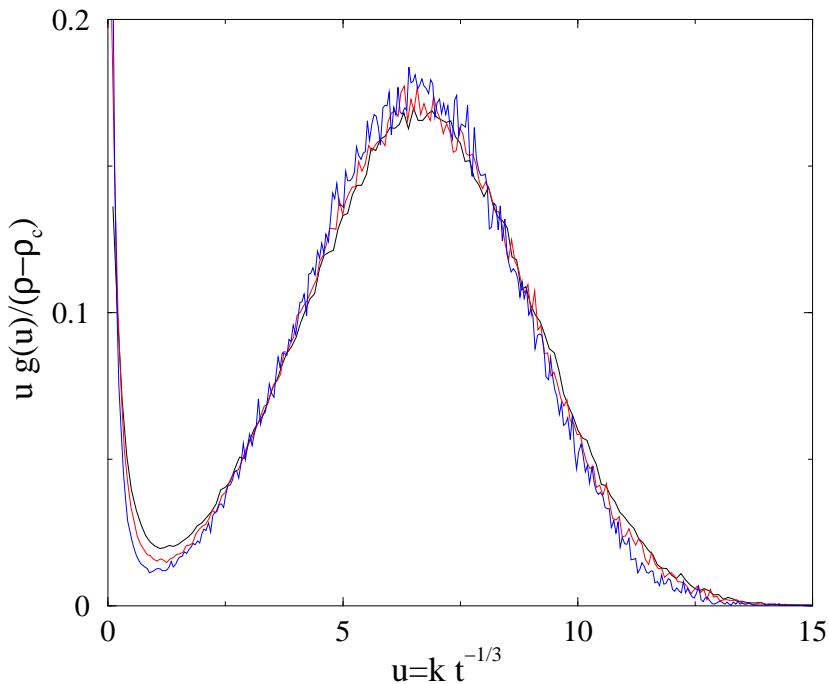}
\caption{\small
One-dimensional system:
normalised scaling function $u g(u)/(\rho-\rho_c)$
in the condensed phase
for times equal to $10^3$, $3\,10^3$,
and $10^4$.
($b=4$, $\rho=20\rho_c=10$.)
}
\label{f3}
\end{center}
\end{figure}
\begin{figure}[htb]
\begin{center}
\includegraphics[angle=0,width=.7\linewidth]{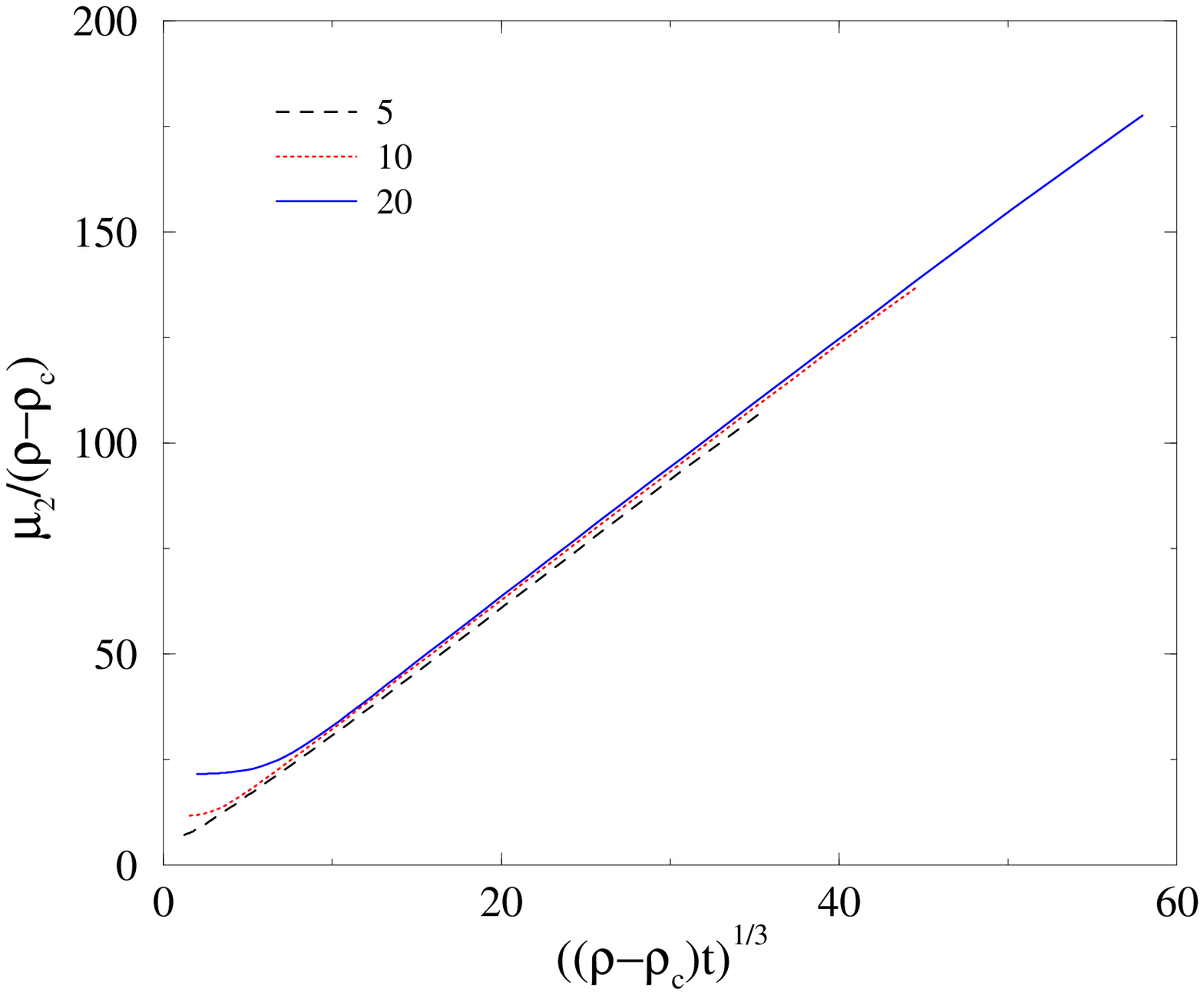}
\caption{\small
One-dimensional system: rescaled mean squared occupation
$\mu_2/(\rho-\rho_c)$
in the condensed phase, for three values of the density:
$\rho=5,10,20$, from bottom to top.
($b=4$.)
}
\label{f4}
\end{center}
\end{figure}

We checked that the observed value of the dynamical exponent
$z=3$ was robust to changes in $b$.
On the other hand, while for the mean-field case the scaling function
$g(u)/(\rho-\rho_c)$
was independent of the density $\rho$, in the one-dimensional case
numerical simulations show dependence in the density.
However this dependence can be absorbed in a rescaling of time:
\begin{equation}
t\longrightarrow (\rho-\rho_c)t,
\label{rescal}
\end{equation}
as demonstrated by figure~\ref{f4}.
It depicts a plot of
 $\mu _{2}/(\rho-\rho_c)$ against $((\rho-\rho_c)t)^{1/3}$,
for $\rho=5,10,20$.
In the linear part of the curves,  the slopes are
all approximatively equal to 3,
in very good agreement with
the measured numerical values of the
integral in the right side of (\ref{k2_cond}),
divided by $(\rho-\rho_c)^{4/3}$.
Coming back to figure~\ref{f3}, plotting
$ug(u)(\rho-\rho_c)^{-2/3}$
against $k((\rho-\rho_c)t)^{-1/3}$
for several densities leads to data collapse,
confirming the rescaling of time
by the density.

We then determined the characteristics of regime I ($k$ fixed, $t$ large),
for  $b=4$ and $\rho=10$.
We first observed that eqs.(\ref{dbart}), (\ref{fkreg1}) and (\ref{vk}) still
hold. This was checked in a number of ways. For instance one can plot ($%
f_{0}/f_{0,\eq}-1)^{-1}$ and $(\bar{d}_{t}-1)^{-1}$against $\mu _{2}$,
yielding straight lines of slopes respectively equal to $v_{0}\,I\simeq -136$
and $A\,I\simeq 59$.
This leads to $A\simeq 0.97$, and $v_{0}\simeq -2.2$.
These values are consistent with the relation (\ref{rs2_cond}), since the
measured value of the integral $\int \d u g(u)\simeq 1.7$.
The amplitude $A$ also depends on $\rho$,
with values equal respectively to
$1.25$, $0.97$, and $0.76$ (for $\rho=5,10,20$),
which, when multiplied by $(\rho-\rho_c)^{1/3}$,
consistently with (\ref{rescal}),
collapse to a value $\simeq 2$.

\subsection{Criticality ($\rho =\rho _{c}$)}

\begin{figure}[htb]
\begin{center}
\includegraphics[angle=0,width=.7\linewidth]{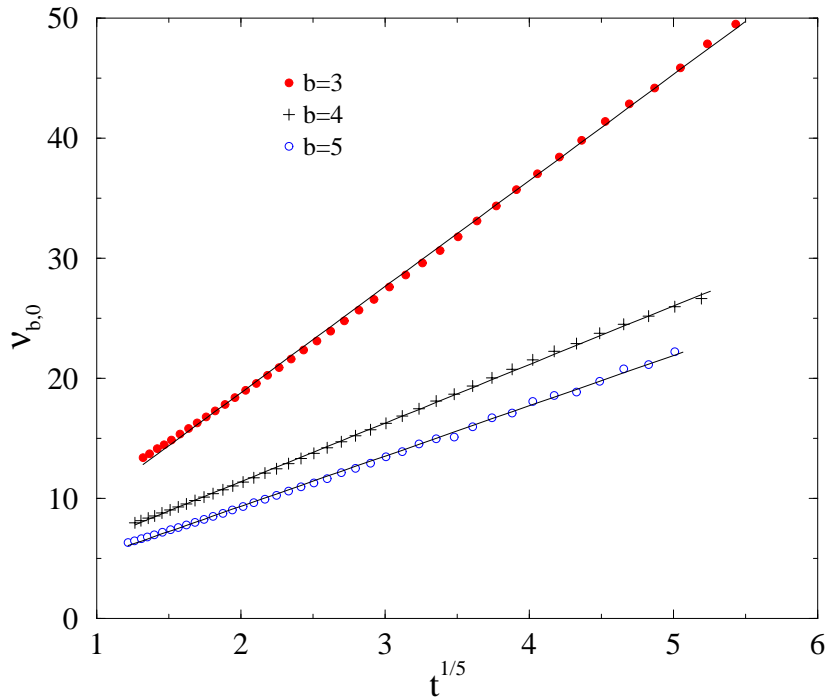}
\caption{\small
One-dimensional system: plot of $\nu_{b,0}$
against $t^{1/5}$ at criticality, for
$b=3, 4, 5$, with regression lines.
($\rho=\rho_c=1/(b-2)$.)
}
\label{f5}
\end{center}
\end{figure}

\begin{figure}[htb]
\begin{center}
\includegraphics[angle=0,width=.7\linewidth]{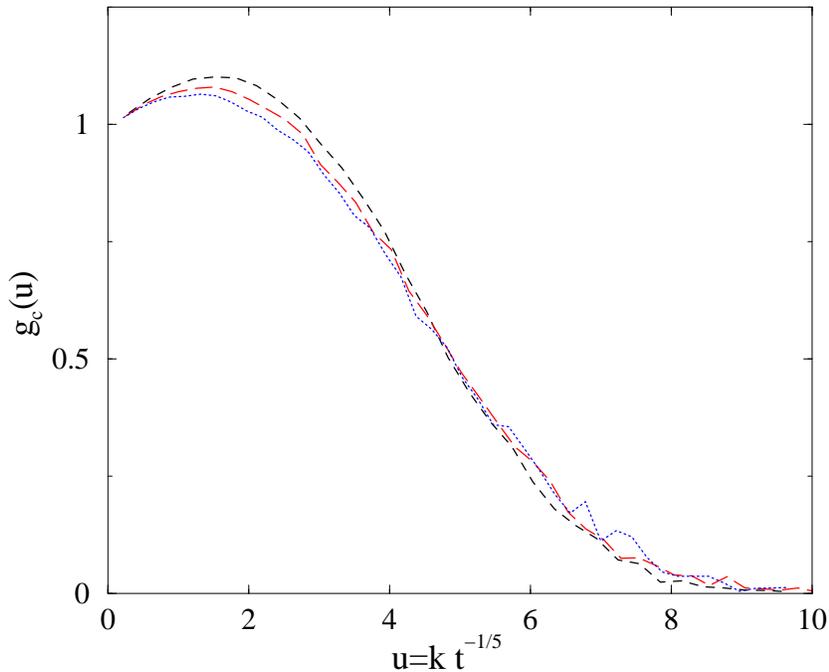}
\caption{\small
One-dimensional system: plot of the critical
scaling function $g_c(u)$
for three different times ($800$, $2000$, and $4000$).\
($b=4$, $\rho=\rho_c=0.5$.)
}
\label{f6}
\end{center}
\end{figure}

The critical case is harder to analyse numerically,
because of the presence of large
fluctuations in the system, and because the asymptotic regime
is longer to attain.
Guided by the mean-field analysis, we expect the dynamics to exhibit two
scales of time, corresponding to regime I and II respectively. In the
latter, let us assume the scaling behaviour, inspired from
(\ref{fkscal_crit}),
\[
f_{k}(t)\approx f_{k,\eq}\,g_c(u)\qquad u=k\,t^{-1/z_{c}}, 
\]
where $z_{c}$ is the critical dynamical exponent.
In order to determine the
value of this exponent, we measured
\begin{equation}
\nu _{b,n}=\sum_{k=0}^{\infty }\frac{f_{k}(t)}{f_{k,\eq}}k^n\sim
t^{(1/z_{c})(n-1)}\int_{0}^{\infty }\d u\,g_c(u),
\label{mub}
\end{equation}
for $n=0$ and $n=1$, and where the notation reminds the fact that,
at large values of $k$, $f_{k,\eq}\sim k^{-b}$.
Figure~\ref{f5} depicts a plot of $\nu_{b,0}$  for $b=3,4,5$
against $t^{1/5}$, which
points towards $z_{c}=5$.
Actually the best regression fit is obtained for $z_c^{-1}\simeq0.215$.
We checked that $\nu_{b,1}\sim(\nu_{b,0})^2$.
The ratio of these two quantities plotted against time gives
an indication of how the asymptotic regime is attained.
Figure~\ref{f6} depicts a plot of the
critical scaling function $g_c(u)$, against the scaling
variable $u=k\,t^{-1/5}$,
for three different times ($800$, $2000$, and $4000$).

In regime I the sum rules (\ref{sumrule1}) and
(\ref{sumrule2}) yield, along the same lines as in section 3.2 (see
eqs. (\ref{v0_plus_aroc}) and (\ref{v0roc})),
\begin{equation}
\omega =\frac{b-2}{z_{c}},  \label{omega1D}
\end{equation}
which generalizes (\ref{omega}).

\subsection{Universality}

As for the mean-field case, we compared the numerical results obtained with
the choice of rates (\ref{dk}), (\ref{other1}), and (\ref{other2}).
For instance, in the condensed phase, consider
the mean squared occupation $\mu _{2}$, defined by  equation~(\ref{mu2}).
The global appearance of the set
of three curves thus obtained
is very similar to the corresponding set of curves obtained
in the mean-field case, for the same choice of rates.
For the latter, one
observes weak discrepancies between the curves at shorter times, which
however disappear at very large times, say $10^{5}$, obtained by numerical
integration of the master equation (\ref{master}).

The same analysis can be done at criticality, for instance
by considering the quantities $\nu_{b,0}$ and $\nu_{b,1}$
defined in (\ref{mub}), with the same conclusions.
In particular the ratio  $\nu_{b,1}/(\nu_{b,0})^2$
converges at large times to a universal amplitude ratio only dependent on $b$.

We are thus lead to infer that, for the one-dimensional system,
the scaling functions describing
the asymptotic behaviour of the occupation probabilities $f_k$
are, as for the mean-field system,
universal quantities, independent of the detailed form of
the hopping rate $d_{k}$,
provided that at large values of $k$, $d_{k}\approx 1+b/k$.
However, in contrast with the mean-field case, in the condensed phase
there exists a dependence on the density through a rescaling of time.

\section{Discussion}

\subsection{Universality: a summary, and the role of a bias}

Much emphasis was given in the present work to universal aspects of dynamics.
Numerical simulations demonstrate the existence of scaling
in the one-dimensional system, very akin to the scaling
behaviour present at the mean-field level.
The scaling regime is characterized by scaling functions,
$g$ and $g_c$, in the condensed
phase and at criticality respectively,
which do not depend on
the detailed form of the rate $d_k$,
provided that at large values of the occupation variable, $d_k\approx1+b/k$.
They only depend on the parameter $b$,
and, in the condensed phase, on the density,
in contrast with the mean-field case.
However a rescaling of time by the density, and correspondingly of $g$,
leads to excellent data collapse.
This universality extends to the case of the
mean-field zeta urn model, though the
definition of the model is different from that of the zero-range processes
studied here.
The dynamical exponents $z$ and $z_c$ are universal in a stronger sense
since they neither depend on $b$, nor on the density.
They only depend on the dimensionality
of space,
with the mean-field geometry corresponding to
the limiting case  $d=\infty$.

The observed differences between the values of the dynamical exponents $z$
and $z_{c}$, in the mean-field geometry on one hand, and in the
one-dimensional case on the other hand, give a measure of the role of
the diffusion of particles from sites to sites.
Indeed,  besides the fact, independent of the geometry of the system, that
the occupation numbers $N_{i}$ are fluctuating variables
by the very definition of the process,
for the one-dimensional system a new phenomenon appears, which
is the spatial diffusion of particles along the line.
This induces correlations between neighbouring boxes (sites).

\begin{figure}[htb]
\begin{center}
\includegraphics[angle=0,width=.4\linewidth]{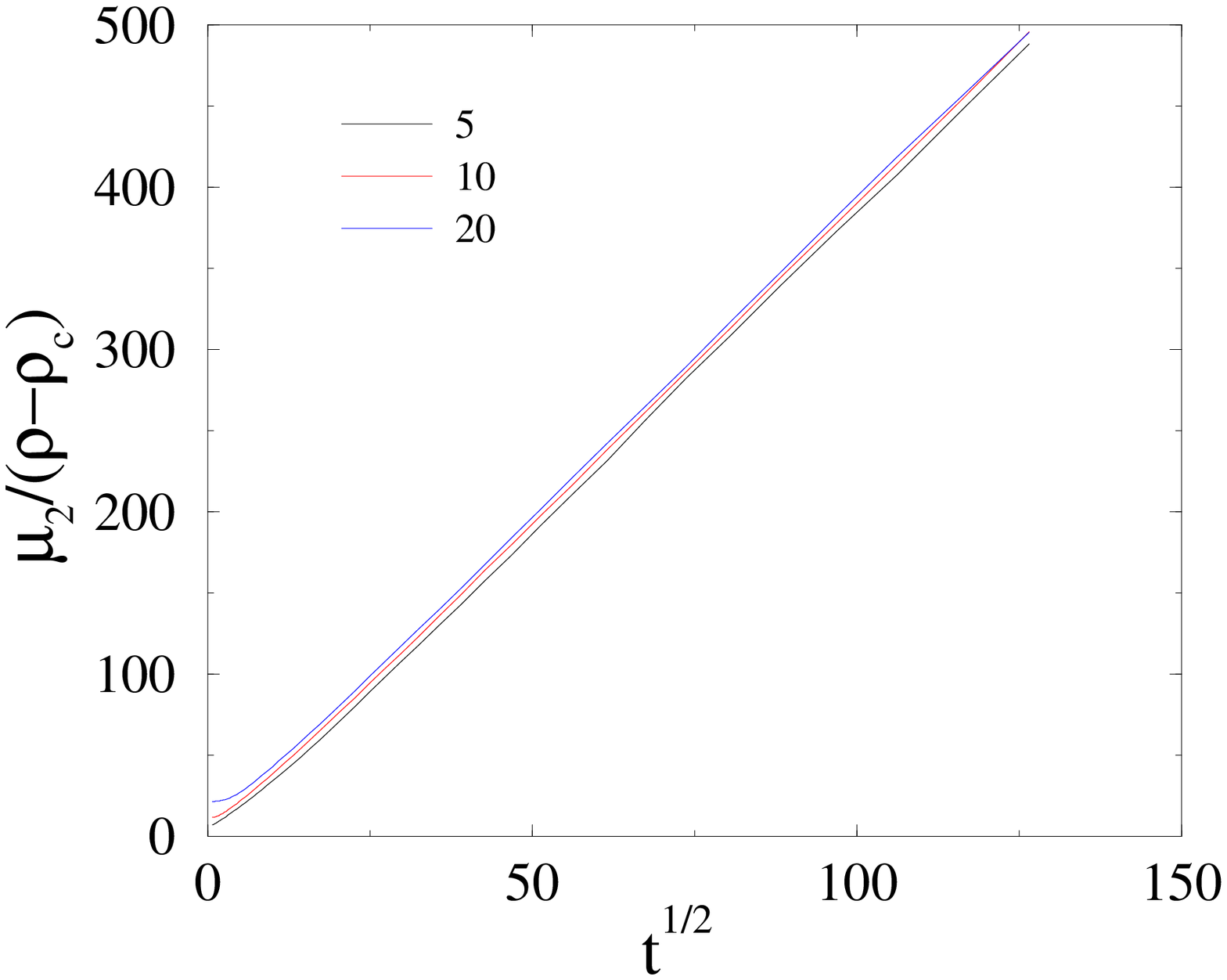}
\includegraphics[angle=0,width=.4\linewidth]{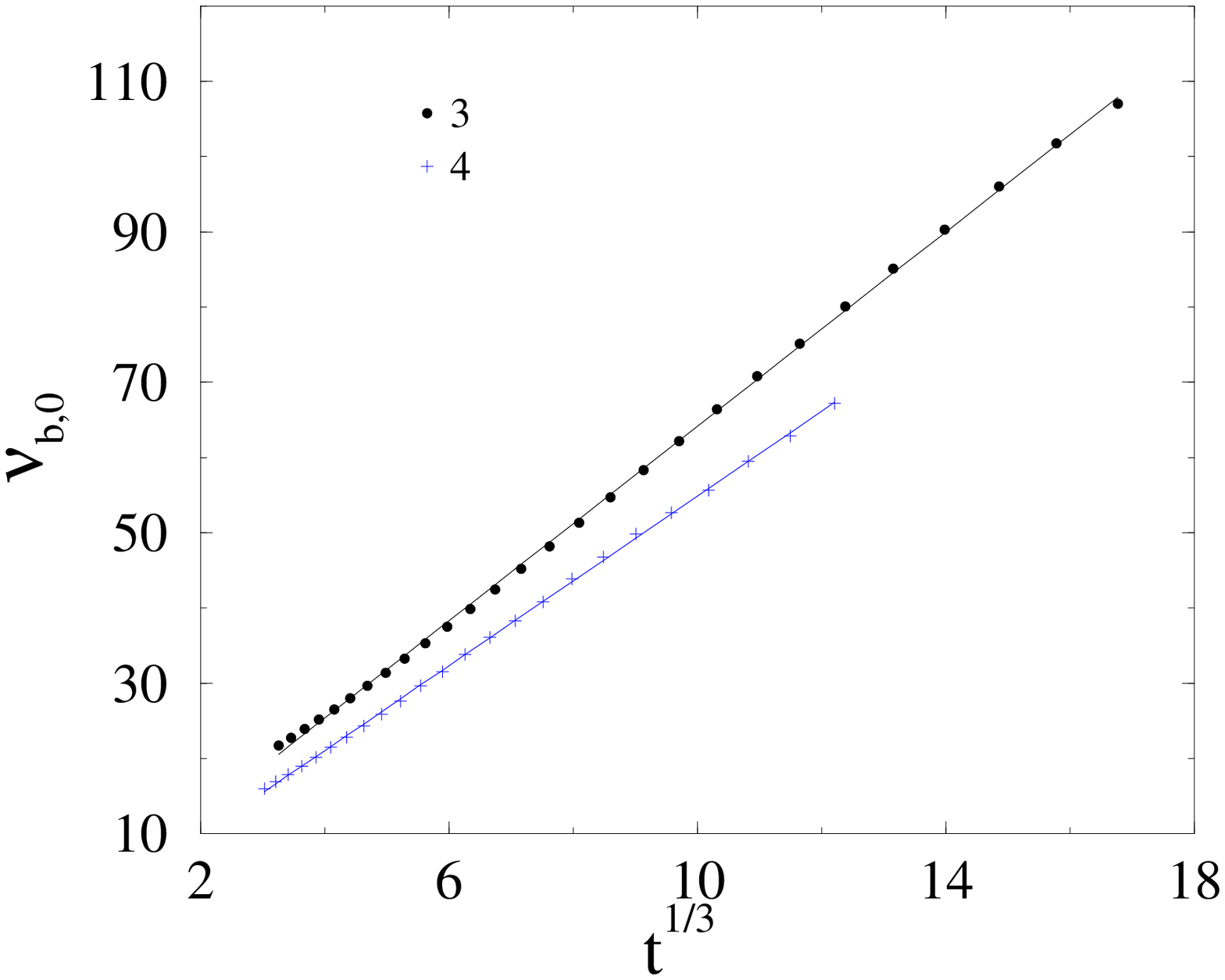}
\caption{\small
One-dimensional system: asymmetric ZRP ($p=1$).\\
Left:
rescaled mean squared occupation
$\mu_2/(\rho-\rho_c)$ against $t^{1/2}$
in the condensed phase, for
$\rho=5,10,20$, from bottom to top. ($b=4$.)\\
Right: $\nu_{b,0}$ against $t^{1/3}$ at criticality, for
$b=3,4$, with regression lines.
($\rho=\rho_c=1/(b-2)$.)
}
\label{f7}
\end{center}
\end{figure}

To gain more understanding of this point, we introduced a bias in the move
of a particle out of a box.
A particle now hops to the right with probability $p$
and to the left with probability $q=1-p$.\
This does not change the stationary state.
For any positive value of the
bias $p-q$ we found $z=2$, and $z_{c}\simeq3$ (see figure \ref{f7}).
Therefore the symmetric and asymmetric one-dimensional ZRP
belong to two different universality classes.
In the extreme case where $p=1$, corresponding to
the totally asymmetric ZRP,
there is no diffusion left in the motion of a particle
along the line.
One could therefore expect that
the totally asymmetric ZRP should resemble,
at least to some extent, to the mean-field ZRP.
It indeed turns out that, in the condensed phase,
the scaling function $g(u)$
of the former
is quantitatively very close
to the scaling function of the latter.
Moreover, for $p=1$, the universality of
the scaling function $g(u)$ with respect to density is restored,
without rescaling of time.
(This point was also checked for the value $p=3/4$.)
However the fact that, seemingly, $z_c\simeq3$
is a caveat that the influence of a bias is more subtle.
As a first step, it would be interesting to
have a better understanding of the role of the bias
on the decay of correlations.

A summary of the different cases studied in this work is presented
in table~\ref{table}.

\begin{table}
\caption{\label{table}Values of the dynamical exponents
for three classes of universality:
mean-field geometry ($d=\infty$),
one-dimensional asymmetric ($d=1,p\neq\frac{1}{2}$),
one-dimensional symmetric ($d=1,p=\frac{1}{2}$).}
\begin{indented}
\item[]\begin{tabular}{llll}
\br
&$d=\infty$&$d=1,p\neq\frac{1}{2}$&$d=1,p=\frac{1}{2}$\\
\mr
$z$&2&2&3\\
$z_c$&2&$\simeq3$&$\simeq5$\\
\br
\end{tabular}
\end{indented}
\end{table}

\subsection{Future work}

Finding an analytical argument in favour of the values of the
exponents appearing in the table is the most immediate challenge.
Other immediate points of interest are the determination of the
upper critical dimension for the zero-range processes under study,
or the explanation of
the role of the density in the one-dimensional
condensed phase.

The value $z=3$ is reminiscent of the value of the dynamical exponent
for the growth of domains in the  dynamics of an  Ising chain
under Kawasaki dynamics, in
the scaling regime of low temperature \cite{cks}.
It is interesting to note that, for the Kawasaki chain,
adding a bias induces a change in the dynamic exponent
from $z=3$ to $z=2$ ~\cite{cornell,spirin}.
Moreover it is simple to see that a mean-field version of
the model leads to $z=2$.
Think to the low-temperature regime, where the detachment
of a spin from a domain is a rare event.
Then whenever such an event occurs, one of the spins which are released
(e.g. the ($+$) spin) is immediately transferred
to a random location inside a domain of the opposite sign.
By diffusion it will shortly stick to the nearest domain of its sign,
thus increasing the size of this domain by one unit,
to the expense of the initial domain, far away in the system.
It is clear from this analysis that the mechanism leading to the
value $z=3$ in the low-temperature coarsening regime of the
Kawasaki chain~\cite{cks} no longer works, hence $z=2$.
Other mean-field versions in the same spirit can be invented,
with the same result regarding the value of $z$.

Trying to draw a closer analogy, beyond these simple observations,
between the dynamics of
the Ising chain under Kawasaki dynamics and the dynamics
of the one-dimensional ZRP
deserves further analysis.

Another point of important interest concerns non-stationary
aspects of the dynamics.
Non stationarity is demonstrated by the behaviour
of two-time quantities such as two-time autocorrelation,
response and fluctuation-dissipation ratio.
At the mean-field level, guided once more by the
analysis of the dynamics of the zeta urn model \cite{zeta2},
we can show that the formalism developed for this model
holds for the ZRP studied in the present
work.
This, as well as the study of the behaviour of two-time quantities
for the one-dimensional ZRP, will be the subject of another
work~\cite{cg}.

The last challenging point concerns the possible relevance
of the present work for
the description of the class of driven particle systems
exhibiting a condensation transition,
such as those studied in \cite{wis2}.
The question posed is whether these particle systems can be
described
\emph{during their temporal evolution},
and not only at stationarity,
in terms of the zero-range processes studied in the present work
\cite{ustocome}.
As a first step one can use the present study
as a template, and ask the following questions~\cite{wis2}.
In the condensed phase, or at criticality,
is there a scaling regime,
and if so, what are the values of the dynamical exponents?
What is the role of the density?
What is the influence of a bias?

\ack
I thank the Einstein Center for support, and the Weizmann Institute,
where part of this work was done, for its warm hospitality.
I am grateful to Y. Kafri,
E. Levine, D. Mukamel, and J. T\"{o}r\"{o}k for interesting discussions on
driven particle systems, and to J.-M. Luck for many fruitful discussions
on urn models.

\end{document}